# Wavelength-commensurate anatase $TiO_2$ particles for robust and functional Mie resonances across the visible and near infrared

Pedro Tartaj,* Yurena Luengo, Pedro Moronta, Luisina Forzani, Alvaro Blanco, and Cefe López†

Instituto de Ciencia de Materiales de Madrid, (ICMM) calle Sor Juana Inés de la Cruz 3, 28049 Madrid, Spain. Consejo Superior de Investigaciones Científicas (CSIC)
E-mail: *p.tartaj@csic.es; †c.lopez@csic.es

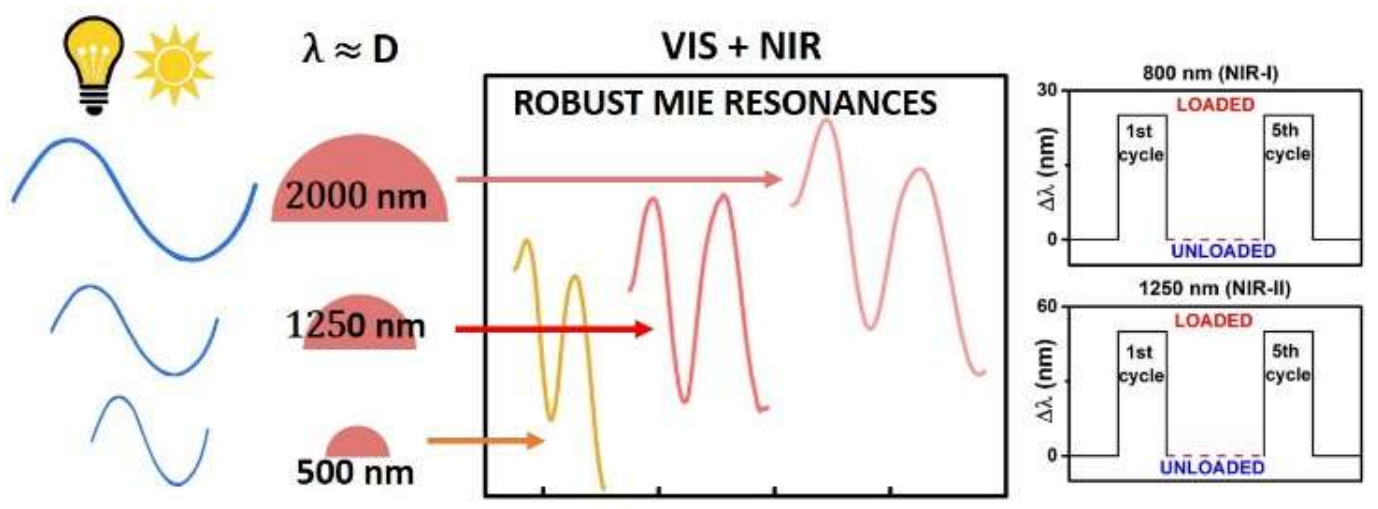

Earth-abundant materials exhibiting Mie resonances across the visible and near-infrared offer opportunities for efficient and sustainable sensing, thermal regulation, and sunlight harvesting. For anatase $TiO_2$, a broadband optical and abundant material, Mie calculations indicate that robust resonances require size tunability and monodispersity (standard deviation, ≲ 5%) over an extended range (0.5-2 μm) not yet experimentally covered, while maintaining a refractive index above 2 to ensure optical contrast, for example with biomolecules. Here, we demonstrate that a simple UV-assisted thermal hydrolysis route yields anatase particles that meet all these criteria and remain aqueous-processable. Consequently, the materials display intense and modulable Mie resonances across the visible/near-infrared regions (including biological windows), outperforming previous results that were limited by size. Strong optical resonances combined with ambient-temperature processability enable robust, broadband, label-free detection of transparent biomolecules. Our insights advance in precise synthesis and Mie-based photonics of an earth-abundant material, and indicate the potential for cost-effective and rapid-on-demand integration via printing technologies, further enhanced by anatase's biocompatibility and photochemical properties.

## Introduction

The development of structured materials whose characteristic dimensions are commensurate with the wavelength of the relevant radiation, enables the emergence of Mie-type resonances that has become a valuable tool for improving efficiency in multiple applications.[1–8] Special interest centres on dielectric materials with high refractive index ($n$) sustaining pronounced and modulable Mie resonances across the visible (VIS) and near-infrared (NIR) frequencies (i.e. solar spectrum). While silicon suffers from strong absorption in VIS frequencies, $TiO_2$ emerges as a viable broadband option, owing to its relatively wide band gap (≈ 3.0 eV) and high $n$ (>2.0). As the second most abundant transition metal oxide after iron oxide, $TiO_2$ is characterized by its chemical stability, low environmental impact, biocompatibility and photochemical properties.[9,10] Thus, any significant advancement involving this material holds strong promise for scalable integration into multifunctional devices.

Unsurprisingly, the past years have witnessed a surge of interest in the fabrication of $TiO_2$-based metasurfaces.[11–17] While metasurfaces offer exceptional sensitivity and specificity, their implementation often relies on advanced and cost-intensive fabrication techniques, restricting their use to high-value-added applications. Enabled by significant progress in printing technologies,[18,19] alternative approaches based on colloidal materials are regaining momentum for rapid-on-demand and low-cost implementations.[20–22]

Our interest lies in developing aqueous-processable, monodisperse anatase $TiO_2$ particles capable of sustaining intense and modulable Mie resonances across the VIS/NIR frequencies, and in demonstrating the functional relevance of these resonances. While amorphous $TiO_2$ may be considered a potential optical alternative to the crystalline polymorphs, concerns regarding chemical stability, density of defects, and impurity levels render its crystalline counterparts more suitable. Regarding particle size, Mie theory calculations indicate that strong resonances spanning the VIS/NIR range require a collection of particles within a size window of 0.5 to 2 μm (**Figure 1**).

Robust and intense Mie resonances require strict size monodispersity, which in this study is measured through the relative standard deviation (SD). Experimental work on monodisperse particles of various materials, including $TiO_2$, suggests that SD should remain below approximately 5% to ensure optical uniformity, resonance intensity, and robustness.[21,23–25]

Dr. Pedro Tartaj, Dr. Yurena Luengo, Dr. Pedro Moronta, Dr. Luisina Forzani, Dr. Alvaro Blanco, Prof. C. López, Instituto de Ciencia de Materiales de Madrid, (ICMM) calle Sor Juana Inés de la Cruz 3, 28049 Madrid, Spain. Consejo Superior de Investigaciones Científicas (CSIC)
Email: * ptartaj@icmm.csic.es; †c.lopez@csic.es
ORCID: (PT) https://orcid.org/ 0000-0001-7323-1545; (YL) https://orcid.org/0000-0002-3780-8527; (PM) https://orcid.org/0009-0008-2735-1544; (LF) https://orcid.org/0000-0002-8242-7993; (AB) https://orcid.org/0000-0002-0015-4873; (CL) https://orcid.org/0000-0001-5635-4463

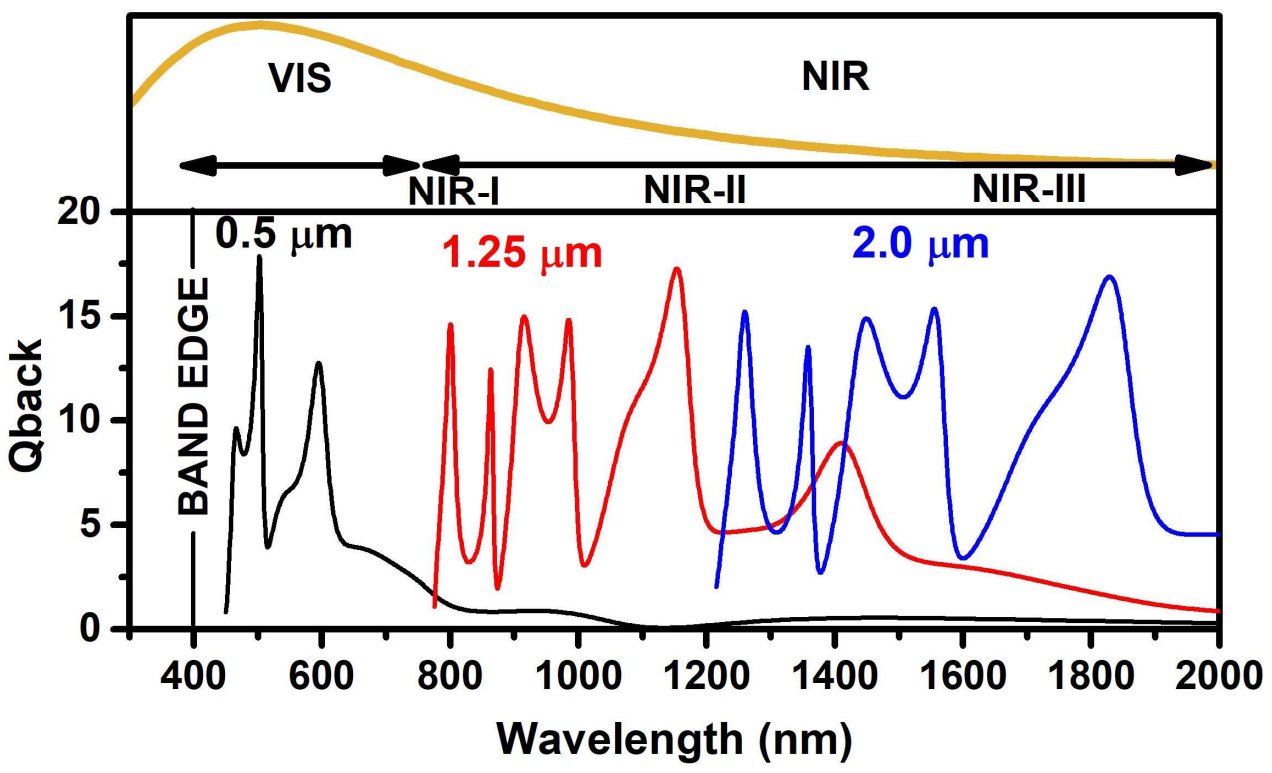

**Figure 1 |** Backscattering coefficient ($Q_{back}$) as a function of wavelength for three different particle sizes (0.5, 1.25 and 2 μm) and *n* = 2.1. Only the strongest and most robust resonances are displayed for each particle size. Size-tunability from 0.5 to 2 μm allows resonances modulation across the VIS/NIR regions, including the biological windows (NIR-I, -II, -III) and a large fraction of the solar spectrum (orange line).

Certain Mie-based applications such as colour filters and photocatalysis involving $TiO_2$ doped with carbon or noble metals can tolerate moderate polydispersity (SD between 5% and 10%).[26–28] However, recent findings indicate that, as might be expected, these systems also benefit from SD ≲ 5% (i.e. more intense resonances).[25] Intense Mie resonances therefore demand SD ≲ 5%, yet the accessible spectral range remains constrained by the particle size employed (≈ 0.5 μm), which the authors of some of those studies (photonic glasses) acknowledge as a bottleneck for achieving spectral Mie tunability across a more extended range.[23] In fact, experimental demonstrations of strong Mie resonances beyond 0.5 μm rely on single-particle scattering measurements rather than on collective effects, due to constrain in uniformity (single-particles are selected from broad size distributions).[29–32]

As mentioned earlier, simulations in Figure 1 clearly show that pronounced and tuneable Mie resonances require a library of monodisperse (SD ≲ 5%) anatase particles spanning a size range from approximately 0.5 to 2 μm. Crystallinity is commonly developed through thermal annealing of monodisperse amorphous particles, which typically leads to a notable contraction in size. Consequently, larger amorphous particles ranging from ≈ 0.8 to 3 μm are required as necessary precursors for achieving the anatase target size.

Extending the uniformity and size-tunability into the demanded range, while keeping the refractive index above 2 remains a significant challenge, despite extensive literature on spherical amorphous/anatase/rutile $TiO_2$ particles (see **Table S1** in supporting information for a summary of previous results).[23,25–28,33–50] Electrostatic strategies have proven effective only at small particle sizes (≈ 300 nm),[38,39,41] while steric stabilization methods or a combination of both stabilizations approaches are routinely employed to achieve higher degrees of monodispersity at larger sizes. However, despite varying degrees of success, these approaches are typically limited to particle sizes below ~1 μm, with the required size-tunability constrained to a narrow window and/or excess porosity that lowers the effective refractive index below 2 (Table S1 in supporting).[23,27,33,37,41,44–46,50]

Our interest lies in methods that make use of the coordination capabilities of ethylene glycol (EG) to produce size-tuneable, monodisperse anatase particles in the 0.15-0.5 μm, following thermal annealing of larger amorphous precursors (0.2-0.7 μm).[40,43,47,48] EG is a non-volatile, odourless solvent extensively utilized in chemical synthesis, with the added advantage that it can be derived from renewable biomass sources.[51] Furthermore, the anatase effective refractive index is greater than 2.0, indicating that optically the porosity is acceptable.[40]

Here, we demonstrate that a modified EG-based method enables the extension of both particle uniformity (SD ≲ 5%) and size-tuneability of anatase into the target range (0.5–2 μm) demanded by Mie predictions, while maintaining the anatase effective refractive index above 2. Furthermore, we show that this key advance is achieved without the use of additional surfactants, relying instead on precisely controlled $NH_3/H_2O$ conditions, even though previous reports in other systems observed a high polydispersity when using aqueous $NH_3$. We also demonstrate that monodisperse anatase particles can be rendered aqueous-processable at ambient temperature via a facile UV treatment. This allows ambient-temperature deposition onto portable substrates by simple casting, resulting in devices that sustain strong and tuneable Mie resonances across the VIS–NIR spectrum. Finally, since the process is compatible with thermally sensitive components and substrates, robust label-free Mie-based sensing of transparent biomolecules in different spectral windows, including biological windows, is demonstrated.

## Results and discussion

### Preparation of monodisperse amorphous precursors

The hydrolysis reported here is a modification of the method previously described by Xia and co-workers,[40] further modified to account for the fact that large spheres require stringent control over experimental conditions and reaction parameters (see Methods). Before proceeding, we observed that the resulting (as grown) amorphous particles are mechanically fragile and chemically unstable. They are prone to degradation during scanning electron microscopy (SEM) imaging and even under prolonged storage in standard laboratory conditions. As a result, particular care was taken during data acquisition to ensure reliable interpretation. As a word of caution, other hydrolysis approaches are likely to exhibit similar behaviour; therefore, data acquisition and analysis should always be conducted with special care.

**Figure 2** displays SEM images of five representative amorphous particle populations ranging in size from 0.8 to 3 μm (see **Table S2** in supporting information for reaction parameters). Size-tuneability with monodispersity is thus extended into a range not previously achieved, without the use of additional surfactants and solely through a simple substitution of water with a concentrated 25 wt% aqueous $NH_3$ solution, alongside appropriate adjustments to the remaining experimental parameters and methodology. As described in more detail below, thermal annealing yields anatase particles spanning the 0.5 to 2 μm size range needed for robust and modulable Mie response.

The outcome departs strongly from other $NH_3$-based hydrolysis routes that do not use EG and produce highly polydisperse particles, with polydispersity increasing with average size (SD

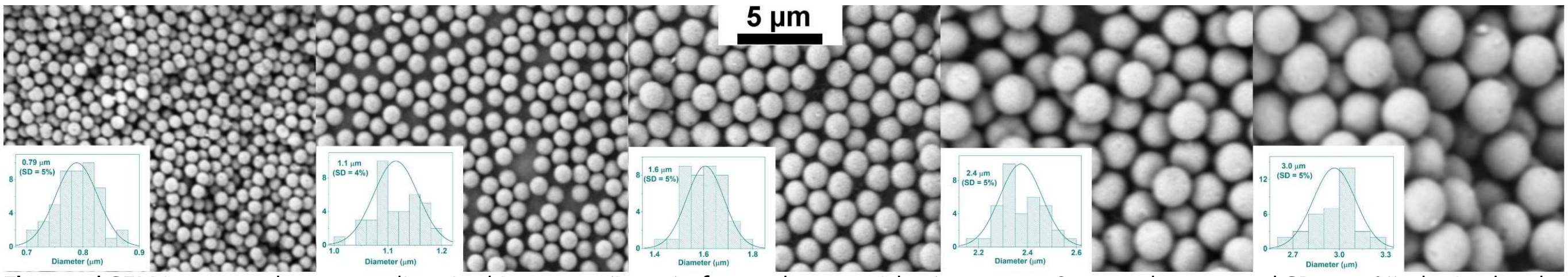

**Figure 2** | SEM images and corresponding size histograms (insets) of amorphous particles (0.79, 1.1, 1.6, 2.4 and 3.0 µm and SD = 4-5%) obtained under different experimental conditions (see Table S2). All images share the same scale bar (5 µm). As noted in the main text, these particles are fragile and chemically unstable, requiring great care during sample preparation and SEM imaging.

≈ 15% and 25% for 500 and 800 nm, respectively).[44] The complexing ability of EG appears to moderate the $NH_3$-catalyzed hydrolysis rate, enabling controlled particle growth via an electrostatic stabilization mechanism previously masked in $NH_3$-only systems. Given that the isoelectric point of anatase is around pH 6, moderating the $NH_3$-driven acceleration of hydrolysis allows growth to proceed under electrostatically stabilized conditions. In essence, regulating the hydrolysis rate renders the system analogous to the $NH_3$-catalyzed hydrolysis of silicon ethoxide in alcohol, which is known to yield highly uniform particles across a broad size range.[52] The proposed role of ammonia as an electrostatic stabilizer is supported by the images presented in **Figure S1**. Under otherwise similar conditions, replacing a concentrated $NH_3$ aqueous solution with only $H_2O$ as the hydrolysing agent leads to extensive particle aggregation, thereby confirming the stabilizing effect of $NH_3$. This stabilization is essential for achieving uniformity at the larger particle sizes reported in this work, where the growth phase plays a dominant role. Basically, if particle–particle aggregation occurs, the diffusion of polymeric species through the liquid phase toward the surfaces of the dispersed particles is no longer possible, making it unfeasible to obtain uniform large particles.

Crucial to the success of the method is the strict control over both the amount of hydrolysing agent and the $NH_3/H_2O$ molar ratio, for which a protocol distinct from the original was developed (see Methods). The formation of these relatively large spheres requires the use of low concentrations of hydrolysing agent. Consequently, all potential sources of external water, including that introduced by solvents, must be minimized to control nucleation rate and batch-to-batch variability. In addition, extreme care must be taken during the washing protocol to ensure reproducibility (see also Methods).

**Figure 3** signals a possible route of experimental conditions free of steep transitions, leading to a broad collection of monodisperse particles of different sizes within the window demanded by Mie simulations (0.5-2 µm for anatase, that is, 0.8-3 µm for the amorphous precursors). Starting from the experimental conditions used to produce the smallest amorphous particles (0.8 µm), gradually reducing the concentration of $NH_3$ (aq) results in an increase in particle size up to a limit where the $NH_3$ (aq) content becomes too low to effectively hydrolyse the butoxide. From this point, we can raise the concentration of EG, the amount of Ti, or even substantially increase $NH_3$, provided that EG, Ti and $NH_3$ are increased together. Thus, up to around 2.5 µm, the synthesis method becomes notably more robust and consistent due to the multiple conditions leading to a specific size. Among different conditions, those using more $NH_3$ offer the advantage of being less sensitive to water contamination. Toward the upper size limit, however, the slope steepens, requiring special care to achieve the largest particle sizes synthesized in this study (3 µm).

## Aqueous-processable monodisperse anatase particles

The conventional route for transforming the amorphous phase into its crystalline counterpart involves annealing at relatively high temperatures. Anatase typically crystallizes first, while higher temperatures are required to induce the rutile phase. During annealing, the spherical morphology is retained; however, a notable size reduction occurs, primarily due to the elimination of organic residues. This was indeed the case for the amorphous particles obtained here; after annealing at 500 ºC they undergo a significant contraction in size **(Figure S2)**. However, during the course of this investigation, we found that the resulting anatases became unsuitable to further processing. This behaviour probably originates from incipient sintering and/or the development of hydrophobic surfaces during the heating at 500 °C.

This work aims to obtain a collection of monodisperse particles within the range demanded by Mie calculations (0.5-2 µm)

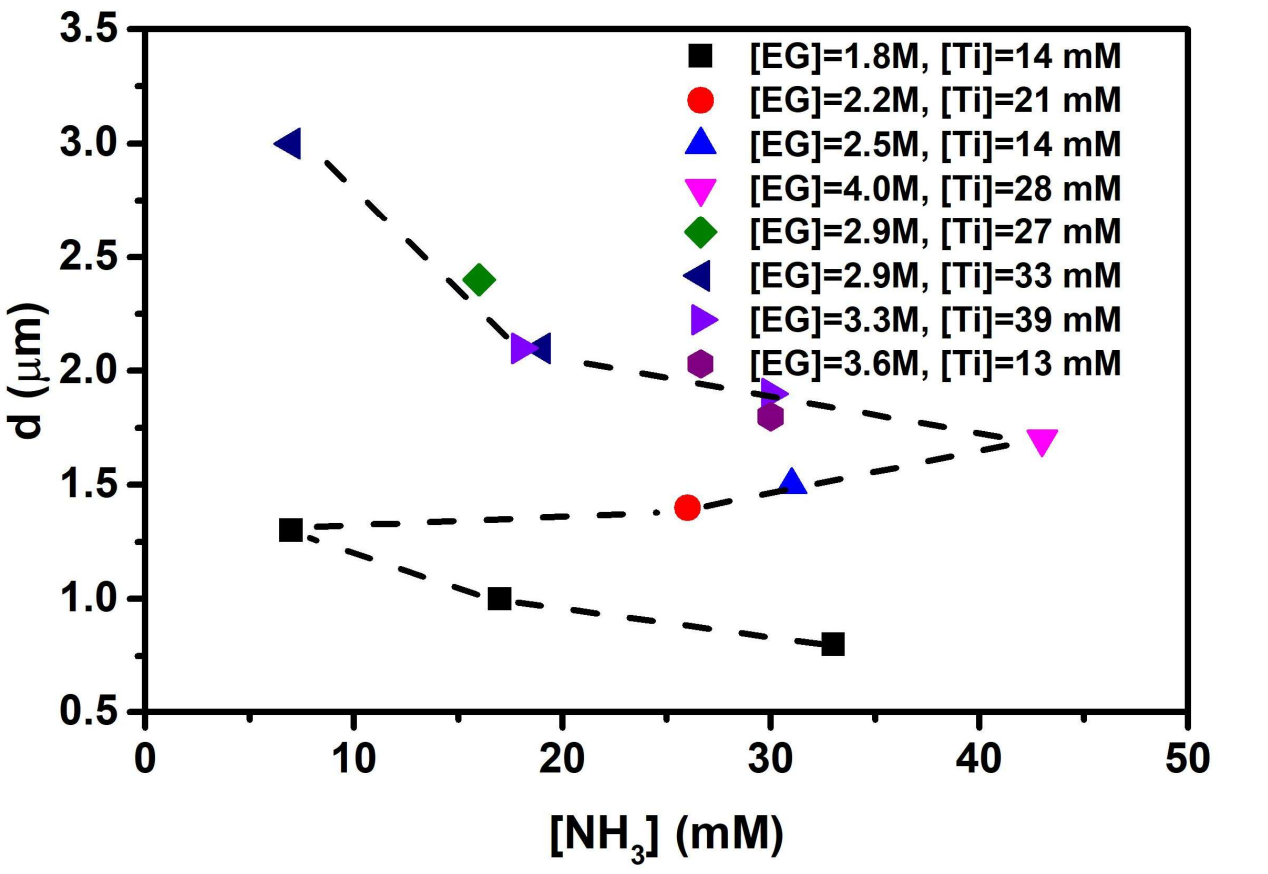

**Figure 3** | Map of experimental conditions defining a smooth pathway toward a robust and extended collection of monodisperse particles, with sizes commensurate with the VIS/NIR frequencies as required by Mie calculations. Particle diameter (d, y-axis) is shown as a function of $NH_3$ concentration (x-axis), with EG and Ti butoxide concentrations indicated in the inset labels. As discussed in the main text, decreasing the concentration of $NH_3$ (aq.) leads to an increase in particle size from 0.8 µm up to a limit beyond which hydrolysis slows down (black squares). From there, the pathway continues through an extended size range where multiple conditions (represented by different symbols) yield comparable diameters, reflecting its robustness. Toward the upper size limit, however, the slope steepens, requiring special care to achieve the largest particle sizes synthesized in this study (3 µm). Note that $NH_3$ is introduced as a 25 wt.% aqueous solution, corresponding to an $NH_3/H_2O$ molar ratio of approximately 1:3 (similar molecular weights).

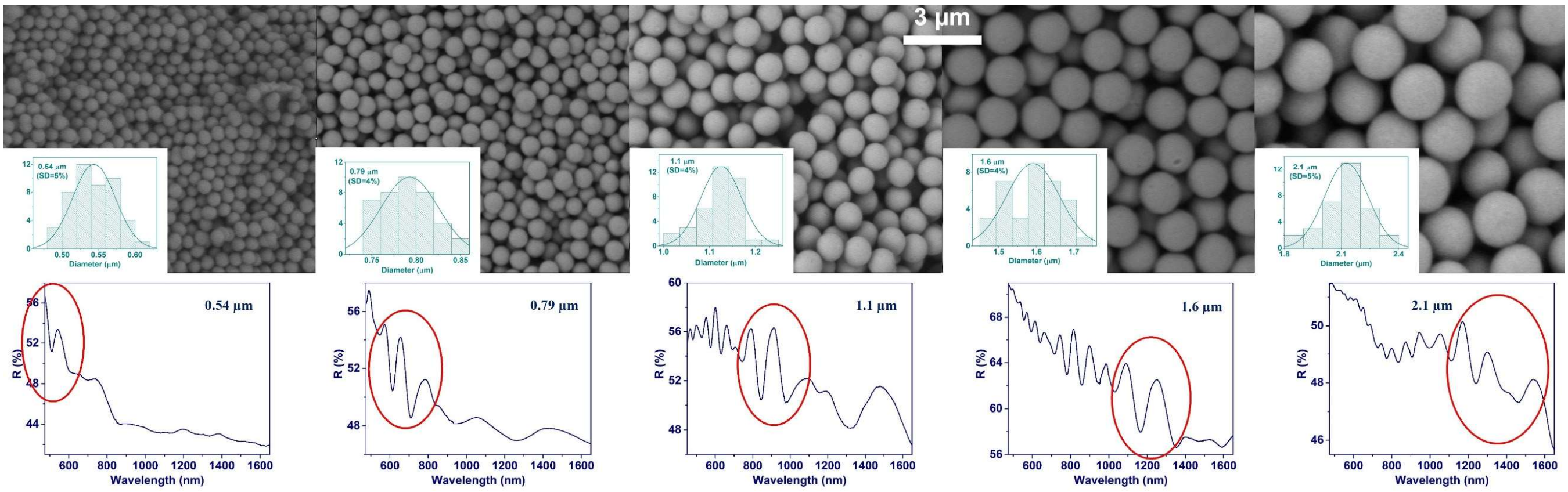

**Figure 4 |** Top horizontal panel: SEM images and corresponding size histograms (insets) of aqueous-processable anatase particles (0.54, 0.79, 1.1, 1.6 and 2.1 µm and SD = 4-5%). Although the overall process led to an approximate 30% reduction in particle size compared to the amorphous precursors, size uniformity was preserved. All images share the same scale bar (3 µm). Low horizontal panel: Total reflectance spectra corresponding to the SEM images shown above. The optical setup limits the wavelength scan to the range of 475 to 1650 nm, which restricts full spectral coverage for the smallest and largest particles (0.54 and 2.1 µm). Nevertheless, size tunability enables the allocation of resonance-rich spectral regions, highlighted with circles, across the VIS/NIR domains. These encircled regions exhibit strong and well-defined signals that are suitable for functional applications such as sensing.

that are processable in aqueous media (under basic pH conditions given that the isoelectric point of anatase, as mentioned earlier, is approximately pH 6). Such compatibility could enable ambient-temperature deposition onto portable substrates and therefore integration with thermally sensitive components, including organic-containing materials and substrates. To achieve aqueous processability, freshly prepared (amorphous) precursor particles were subjected to UV treatment in the presence of small amounts of organic peroxides, facilitating their transformation into the anatase phase at a temperature of 375 °C (see Methods). Elemental analysis confirmed the removal of carbon post-treatment, while X-ray diffraction and infrared spectroscopy verified the formation of anatase (**Figure S3**). Particle size uniformity was preserved throughout the UV treatment and subsequent thermal annealing, as shown in **Figure 4.**

Basically, after the UV treatment, a substantial reduction in residual carbon content, from 15 wt. % to approximately 2–3 wt. % is observed. Furthermore, the elemental hydrogen content remained high (2–3 wt%), suggesting that UV exposure also induced surface hydroxylation, likely via chemisorbed and physisorbed $H_2$, similar to what was observed in UV-treated $TiO_2$ films.[53] The increase hydrophilicity along with the fact that milder temperatures are needed to achieve full crystallization could explain the good processability of the resulting anatases. Further insights into the elimination process are provided in **Figure S4** of the supporting information.

## Mie modulation and label-free optical detection

A brief ultrasonic treatment (standard ultrasonic bath for approximately 5 minutes), followed by gentle shaking, was sufficient to disperse the anatase spheres in diluted aqueous $NH_3$ solutions. As mentioned earlier, the isoelectric point of anatase is near pH 6, which facilitates dispersion under basic conditions. Consequently, anatase can be conveniently cast onto different substrates (widely available glass slides in our case) and exhibit strong and modulable Mie resonances across the visible and NIR spectral regions whose more intense signals can be relocated by changing their size (**Figure 4**). When reflectance is plotted as a function of the dimensionless frequency, $x = 2\pi a/\lambda$, where $a$ is the particle radius, excellent spectral overlap is observed across the five representative samples (**Figure 5**). The observed overlap suggests that the samples share a similar refractive index. Mie calculations for single spheres show reasonable agreement with the experimental data when using a refractive index of $n = 2.1$, previously estimated from Bragg diffraction measurements on anatase photonic crystals composed of 250 nm particles.[40]

The full potential of the method presented here is tested by the ambient-temperature fabrication of a hybrid composite incorporating a biomolecule, with arginine serving as a representative example. Arginine, an amino acid that is optically transparent across the visible and NIR spectral regions, has a refractive index of ≈ 1.5, which is typical of many biological compounds. A concentrated suspension containing both anatase and arginine can be prepared at ambient-temperature and conveniently cast onto glass slides (see Methods).

**Figure 6** presents a comparison of the total reflectance spectra between a substrate containing only anatase and a substrate containing both arginine and anatase of two different particle sizes (1.1 and 1.6 µm). As shown in Figure 4, the 1.1 µm particles exhibit the strongest Mie resonances around 800 nm that includes the first biological window (NIR-I). For the 1.6 µm particles, the strongest resonances occur between 1000 and 1300 nm, which includes the second biological window (NIR-II).

For the 1.1 µm particles, distinct shifts in the Mie resonance position (20 and 25 nm) are evident after incorporating arginine. By selecting a different particle size (1.6 µm), we were able to shift the spectral region for detection toward longer wavelengths (1000–1350 nm). Distinctive Mie resonance shifts of 45 and 50 nm were clearly observed upon incorporating arginine, demonstrating that the spectral range for biomolecule detection can be tuned by particle size selection, making it truly versatile. In contrast, for commercial silica and polystyrene particles, resonances are no longer observed, and even the total reflectance drops to a few percent as corresponds to a refractive index matched layer (refractive index of silica, polystyrene, and biomolecules around 1.5).

Notably, when the composite anatase/arginine is subsequently exposed to a pH 4 solution for 30 minutes, arginine dissolves and the original anatase spectrum, including the resonance positions, remerges (Figure 6). Loading and unloading cyclability studies carried out with the two different particle sizes and at different spectral regions did not show any degradation, reflecting the high chemical stability of anatase (Figure 6). The potential applicability of this system is further supported by the good biocompatibility of $TiO_2$, which has been extensively reported in studies involving amperometric biosensors.[54] Building on these characteristics, the ability to process monodisperse anatase particles at ambient conditions, combined with their responsiveness across multiple spectral regions, opens the door to full integration into low-cost biodevices through printing technologies. Such systems could be used to monitor loading and unloading events in controlled-release patches, for instance, offering a versatile platform for broadband sensing.

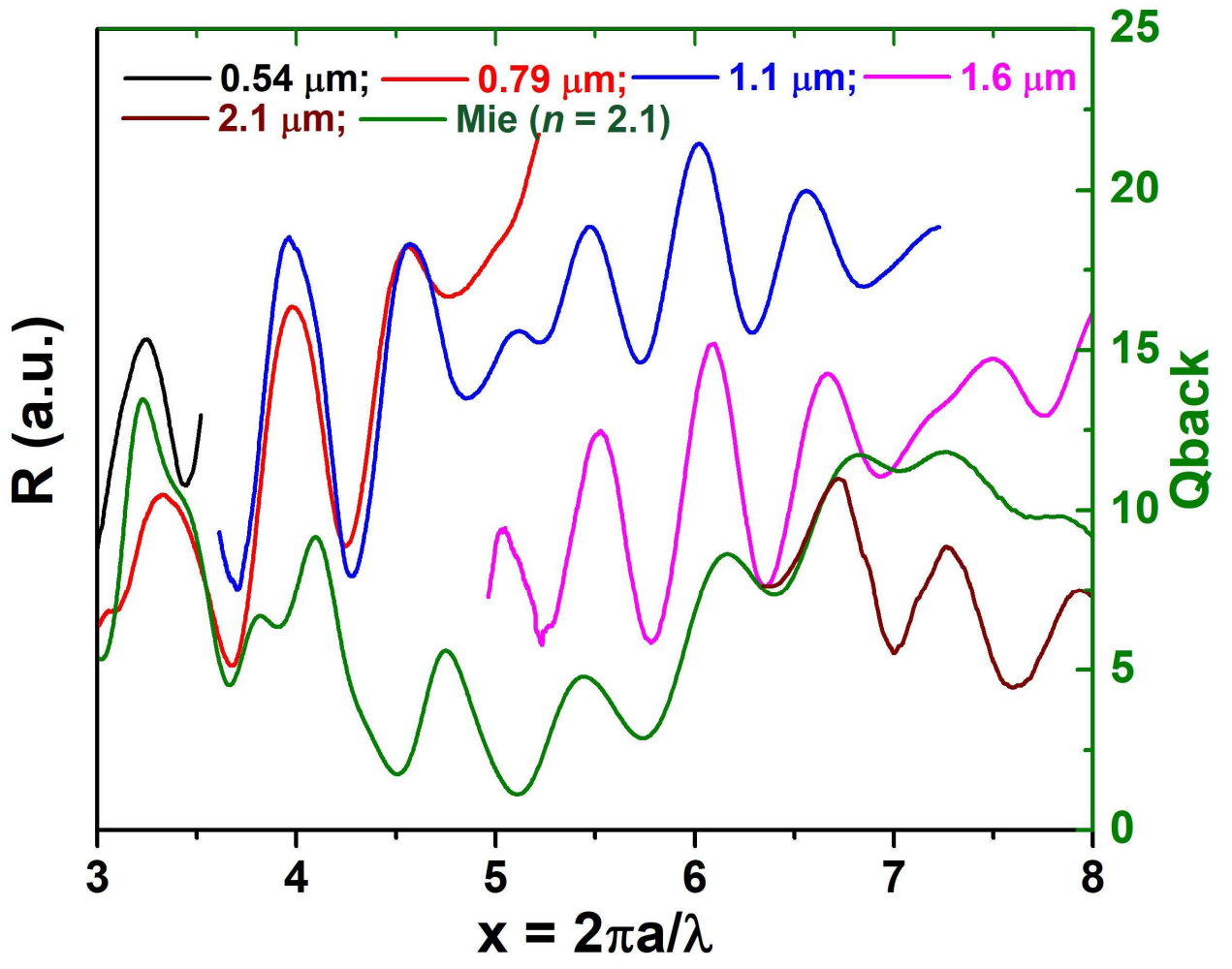

**Figure 5 |** Left vertical axis: Total reflectance spectra as a function of the dimensionless frequency, $x = 2\pi a/\lambda$, where $a$ is the particle radius coming from the SEM characterization. The spectra correspond to the particle diameters shown in Figure 4. To account for the experimental error (SEM calibration and scale bar, counting) in comparing $x$, the diameters were left to float to obtain the best fits (see Methods for details). Overall, the good overlapping of the samples indicates a similar refractive index, $n$, for all samples. Right vertical axis: Mie backscattering coefficient ($Q_{back}$) calculations weighted by a normal size distribution and calculated as a function of the dimensionless frequency, $x$. For the calculations, a value of $n$ = 2.1 was used (see Methods for details).

## Conclusions

In summary, we have demonstrated that monodisperse anatase particles can be synthesized with adjustable sizes within the range of 0.5 to 2 μm, while maintaining a refractive index above 2.0. The dimensions are commensurate with VIS/NIR wavelengths, enabling the emergence of intense and adjustable Mie-type resonances across this spectral domain. Furthermore, we show that this size range is achievable without the use of additional surfactants, solely by replacing water with ammonia as the hydrolysing agent, in combination with the low-impact solvent ethylene glycol. We have shown that crucial to the success of the synthesis is establishing a protocol that accounts for the stringent control required to form large spheres, particularly in terms of reaction parameters and experimental conditions. We demonstrate that monodisperse anatase particles can be rendered processable at ambient temperature in aqueous media via a simple ultraviolet treatment in the presence of low concentrations of organic oxidants. This method enables room-temperature deposition onto portable substrates, producing materials with strong and tuneable Mie resonances across the VIS/NIR range, while remaining compatible with thermally sensitive components. As a result, robust broadband label-free optical detection of biomolecules becomes feasible through measurable shifts in resonance position, including within biologically relevant spectral regions.

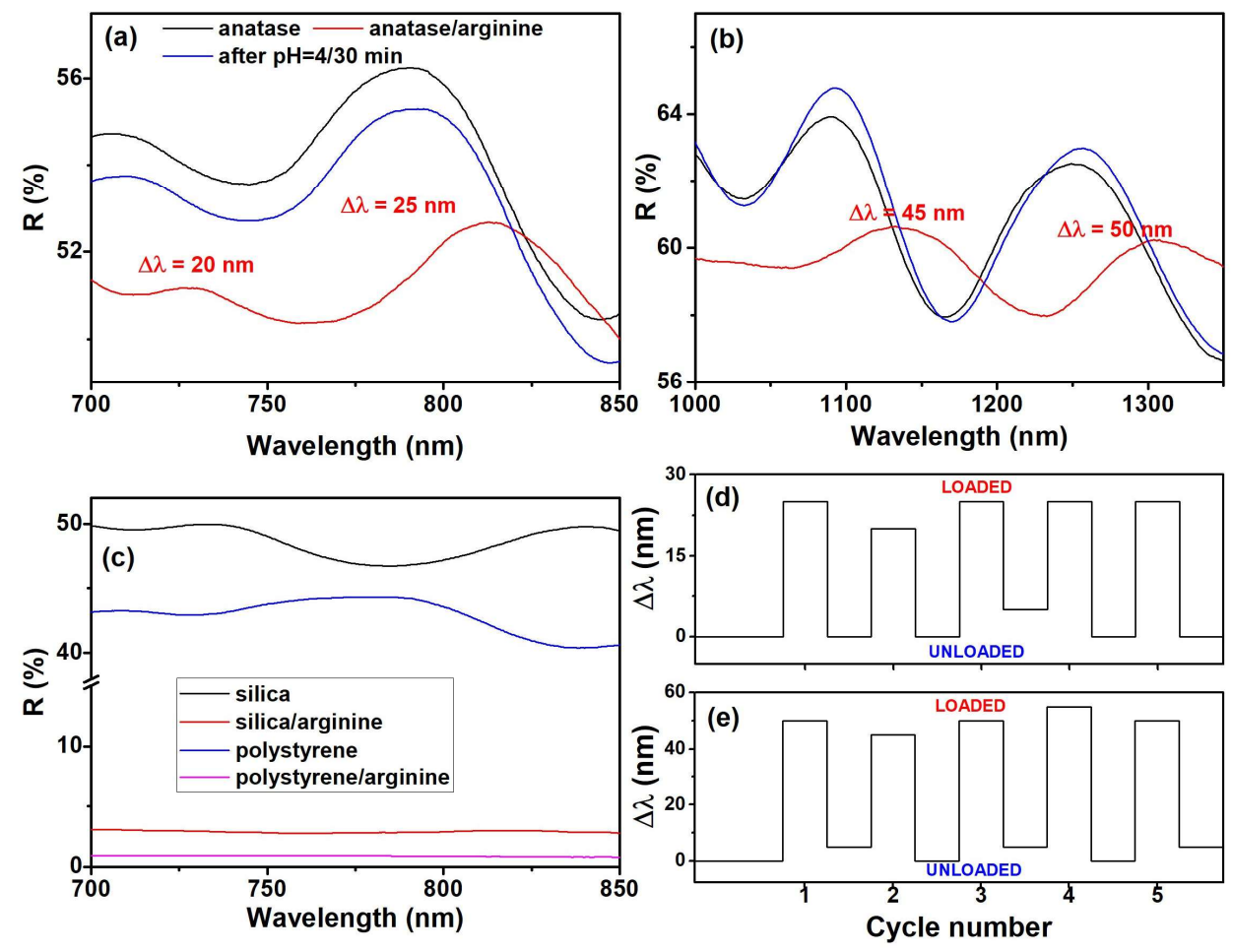

**Figure 6 |** Total reflectance spectra for (a) uniform anatase particles of 1.1 μm before and after mixing with arginine, followed by treatment of the composite in a pH 4 solution for 30 minutes; (b) same procedure applied to 1.6 μm anatase particles; (c) uniform silica and polystyrene particles of 1.5 μm before and after mixing with arginine. The selected wavelength range in this case matches that used for the 1.1 μm anatase particles, compensating for the lower refractive indices of silica and polystyrene. (d) Loading and unloading cyclability studies for the 1.1 μm particles monitored through a resonance peaking at around 800 nm and (e) the 1.6 μm particles monitored through a resonance peaking at around 1250 nm. $\Delta\lambda$ values were rounded to the nearest 5 nm increment. Note: for silica and polystyrene, the y-axis scale was broken to better visualize the Mie resonances. As a result, the reflectance drops after arginine incorporation, from approximately 50 percent to below 3 percent, is more pronounced than it may appear.

## Methods

**Chemical reagents:** Titanium butoxide (97%) and anhydrous ethylene glycol (EG, 99.8%) were sourced from Aldrich and stored under dry conditions. Acetone (EMSURE, nominal $H_2O$ content less than 0.05%) was obtained from SUPELCO. NH3 (aq) (25 wt.%, EMSURE, Emparta), Luperox (TBH70X), and absolute ethanol (Emparta, Supelco) were also used. Monodisperse spherical polystyrene (PS) and silica particles were acquired from Microparticles GmbH.

**Drying of solvents**: As mentioned in the main manuscript a fundamental aspect to assure reproducibility is to dry the commercial solvents. Karl Fischer titration using the kit recommended for acetone was used to estimate the amount of water present in acetone. Typically values around 1000 ppm were estimated for the acetone here used (being around 3000 ppm for other commercial acetones typically labelled as acetone 99.6%). Residual water around 1000 ppm was sufficient to prevent the dissolution of titanium butoxide in acetone via hydrolysis leading to uncontrolled precipitation of amorphous $TiO_2$. To resolve this issue, the acetone was treated with zeolites (30 wt.%). After vigorous shaking, the mixture was left to rest and sediment for about one week. The clear solvent was then filtered, and stored under dry conditions. Following this treatment, water content decreased to below 100 ppm as determined by titration, allowing titanium butoxide to fully dissolve in acetone almost immediately. During this process no condensation to mesityl oxide was observed through UV/VIS spectroscopy. For anhydrous EG, the residual water content was measured at approximately 200 ppm. Attempts to further dry EG using zeolites were unsuccessful, as its high viscosity interfered with effective separation of zeolites leading to rapid aggregation of spherical particles and also to contamination of the final product. A content of 200 ppm was considered thus acceptable given the quantities of EG used in particle synthesis.

**Synthesis of amorphous particles:** In a typical synthesis procedure, the appropriate amount of titanium butoxide was dissolved in 54 mL of acetone and stirred for 10 minutes to avoid prolonged contact of the butoxide with acetone minimizing the formation of a complex (results after 2h of stirring led to similar products). Studies in sol-gel chemistry have shown that ketones, like acetone, form coordination complexes with titanium atoms in alkoxides.[55–57] In a separate vessel, the required volume of $NH_3$ (aq) was mixed with the desired amount of EG (for practical reasons a more concentrated $NH_3$ solution in EG was diluted with EG to the desired value). Note that to further improve consistency and efficiency, we modified the reagent mixing method. Instead of dissolving the titanium alkoxide in ethylene glycol, as in the original protocol,[40] we dissolved it in acetone. As mentioned in the prior section, the absence of residual water in solvents, allows dissolution of the alkoxide in the acetone. Once dried, acetone allows rapid dissolution of the alkoxide, while ammonia is separately dissolved in EG. This adjustment reduces the dissolution time of the alkoxide from over ten hours to just a few minutes and simultaneously enhances stability against hydrolysis.

The $NH_3$–EG and titanium alkoxide–acetone solutions were combined under vigorous stirring, typically by pouring the $NH_3$–EG solution into the titanium alkoxide–acetone mixture contained in a separate vessel. After one minute, the stirring speed was reduced to approximately 200 rpm. Typically, the hydrolysis was stopped after three hours, measured from the moment turbidity first became visually apparent. Importantly, as the visual onset of turbidity is closely related with induction times, we found that monitoring this parameter was useful for the sake of reproducibility. Upon completion, the resulting solid was separated from the reaction mixture via vacuum filtration followed by washing with absolute ethanol to eliminate unreacted alkoxides. This washing protocol was stablished to avoid possible secondary nucleation by using water during the washing procedure (see Figure S5 in supporting information). After washing, the solid was dried at 50 °C. All the reactions were carried out under dry conditions.

**UV treatment:** In a typical experiment, approximately 50 mg of dry amorphous particles were immersed in 20 mL of water and allowed to soak for 2 to 3 hours. The suspension was then vacuum filtered to collect the solid, which was subsequently redispersed in 20 mL of water using ultrasound for about one minute. Next, 100 µL of Luperox were added to the suspension, which was then exposed to ultraviolet (UV) light using a 200 W Xe-Hg lamp operating at a peak wavelength of 365 nm and an intensity of approximately 5 mW/cm$^2$.The suspension was gently stirred at 100 rpm throughout the irradiation. After one hour, the sample was again redispersed using ultrasound for about one minute, followed by the addition of another 100 µL of Luperox and continued UV exposure under weak stirring. This cycle was repeated once more at the two-hour mark, and the overall degradation process concluded after three hours of total irradiation.

**Annealing:** The annealing procedure followed a multistep temperature ramp to reach the final setpoint. For amorphous particles annealed at 500 °C for 5 hours, the samples were first heated to 150 °C at a rate of 5 °C per minute and held for 2 hours. Subsequent heating steps were performed at intervals of 100 °C, specifically at 250 °C, 375 °C, and 450 °C, each maintained for 2 hours, before reaching the final temperature of 500 °C. For UV-treated particles annealed at 375 °C for 10 hours, a similar ramping protocol was followed. However, intermediate steps at 200 °C and 300 °C were included. The temperature sequence consisted of holds at 150 °C, 200 °C, 250 °C, and 300 °C, each for 2 hours.

**Casting and cyclability studies:** In a typical experiment, a suspension was prepared by dispersing the particles at a concentration of 5 mg/mL in 1 M $NH_3$ aqueous solutions. The mixture was placed in polyethylene tubes (the typical ones with green screw caps), vigorously shaken, and subjected to ultrasonic treatment for about 5 min. Once the suspension was homogenized, it was cast onto glass substrates. For composite preparation, a solution of arginine also in 1 M $NH_3$ was prepared separately. This solution was then combined with the particle suspension and shaken in a commercial vortex. The resulting composite mixture was subsequently cast onto glass substrates. After drying, the resulting composite film is suitable for optical characterization via total reflectance spectroscopy. The content was arginine/anatase (50/50 w/w). To eliminate arginine, the glass substrate was carefully immersed in a pH 4 solution following optical characterization. After 30 minutes, the solution was gently replaced with water three times, and the substrate was subsequently removed, dried, and recharacterized optically. For cyclability studies, the entire process was repeated from the beginning up to five times. To enable this, a suspension containing the arginine–anatase mixture was prepared in sufficient quantity. A small portion was deposited onto a glass substrate, dried, and subjected to optical characterization before and after immersion in the pH 4 solution. The remaining larger fraction was deposited onto a glass dish, dried, and exposed to the same pH 4 treatment and washing conditions. From this portion, a fresh anatase dispersion was prepared with the required amount of arginine, marking the beginning of the second cycle. The process was repeated five times. This approach ensured consistent composition and deposition conditions throughout the study.

**Chemical and structural characterization**: The morphology and particle size were examined by transmission electron microscopy (TEM, 2000 FX2, JEOL) and scanning electron microscopy (SEM, FEI Nova NanoSEM 230 microscope with a Schottky field-emission gun). The mean size and the standard deviation (SD) were evaluated from SEM pictures by counting around 50 particles. The uniformity in size was evaluated from the relative SD (SD/mean size). Infrared spectra were recorded using a Bruker IFS 66VS after dilution in KBr (2 wt.%). DTA/TG analysis were carried out in a TA Instrument Q600 under dynamic air (100 mL/min) with a heating ramp 10 °C/min, while elemental analysis of C, H and N was determined in LECO TruSpec Micro. X-ray diffraction patterns were collected from 5 to 70° (2θ) by using a Bruker D8 Advance instrument with CuKα radiation and a SOLX detector operating at 40 kV and 30 mA.

**Optical characterization:** The optical characterization of samples was carried out by total (diffuse + specular) reflectance measurements with an integrating sphere coupled to different compact spectrometers all from Ocean Optics. The spectrometers have different sensitivity and spectral resolution: an Ocean 2000 covered data from 500 to 900 nm, and a NIRQUEST 500, set to high gain mode, covered the 900 to 1650 nm NIR region. Mie backscattering coefficient ($Q_{back}$) calculations for individual spheres were carried using the free Python code developed by Scott Prahl.[58] A function to calculate $Q_{back}$ weighted with a normal distribution was developed and added to the python code. To account for experimental uncertainties such as SEM calibration, scale bar resolution and particle counting, the diameters used to calculate the dimensionless frequency, $x = 2\pi a/\lambda$, were allowed to vary within ±5%, reflecting the system's high sensitivity to small size variations. Mie scattering calculations were performed as a function of $x$, using a refractive index of $n$ = 2.1 for all wavelengths. For the 2.1 µm particles, the best agreement was found at 2.02 µm. For the 1.6 µm particles the best fit was found at 1.58 µm. The 1.1 µm samples showed optimal overlap at 1.15 µm, while the 0.79 µm particles aligned best at 0.83 µm. Finally, for the 0.54 µm particles, the best fit was achieved at an effective diameter of 0.56 µm. Notably, in all cases the refined diameters remained within a ±5% threshold, reinforcing the consistency of the model fitting.

## Supporting Information

Supporting information can be found after the LIST OF REFERENCES

## Acknowledgements

This work was partially funded by the Spanish PID2021-124814NB-C21, PID2024-158675OB-C21, the "Severo Ochoa" Seal of Excellence in R&D (CEX2024-001445-S).

## Data Availability Statement

The data that support the findings of this study are available from the corresponding author upon reasonable request

## Author contributions

P.T., Y.L., P.M., and C.L. performed conceptualization. P.T., A.B, and C.L. performed supervision. P.T., and Y.L., performed the methodology. P.T., Y.L., P.M., L.F., and A.B. performed the investigation. P.T. performed software, formal analysis, data curation, and visualization. P.T., A.B., and C.L. performed funding and resources. P.T., Y.L., and C.L. wrote-original draft. P.T., Y.L., P.M., L.F., A.B., and C.L. wrote, reviewed, and edited.

SUPPORTING INFORMATION FOR

# Wavelength-commensurate anatase $TiO_2$ particles for robust and functional Mie resonances across the visible and near infrared

Pedro Tartaj,* Yurena Luengo, Pedro Moronta, Luisina Forzani, Alvaro Blanco, and Cefe López†

Instituto de Ciencia de Materiales de Madrid, (ICMM) calle Sor Juana Inés de la Cruz 3, 28049 Madrid, Spain. Consejo Superior de Investigaciones Científicas (CSIC)

E-mail: *ptartaj@icmm.csic.es; †c.lopez@csic.es

**Table S1:** Survey of literature data on spherical $TiO_2$ particles limited to sizes between 0.2-3 µm, slightly exceeding the size range of primary interest in this study. The survey is also limited to standard deviations (SD ≲ 20%), except when a relevant particle size within a given study exceeds this threshold. In cases where both amorphous and anatase particles are analysed, only data corresponding to anatase is included. To broaden the scope, the table also lists selected high-surface-area mesoporous materials that are uniform and, in our view, relevant, despite their low effective refractive index ($n_{eff} < 2$) resulting from the porous structure. To facilitate reading, entries are ordered by particle uniformity, from lowest (our requirement) to highest standard deviation. References to studies reporting comparable outcomes using similar methods have been omitted for conciseness, especially when employing a different approach from the one described in our study.

| Size (SD) | Phase | Size Estimation | Method | Source |
|---|---|---|---|---|
| 0.2-0.4 µm (≲ 5%) | Anatase | Given by the authors | Hydrolysis in the presence of ethylene glycol followed by annealing | Jiang et al.,[1] |
| 0.5 µm (≲ 5%) | Anatase | Given by the authors | Hydrolysis in the presence of ethylene glycol and Tween-20 followed by annealing | Yu et al.,[2] |
| 0.25-0.4 µm (≲ 5%) | Anatase | From a size evolution curve assuming error bars represent minimum and maximum observed values | Hydrolysis in the presence of ethylene glycol followed by stirring at 70 C and then annealing | Cheng et al.,[3] |
| 0.5 µm (≲ 5%) | Anatase | Given by the authors | Hydrolysis in the presence of ethylene glycol followed by thermal annealing | Lim et al.,[4] |
| 0.8 µm (≲ 5%) | Anatase | Given by the authors | Hydrolysis with KCl aqueous solution in the presence of hexadecylamine followed by solvothermal treatment and annealing (mesoporous so $n_{eff}$ likely < 2) | Chen et al.,[5] |
| 0.5 µm (≲ 5%) | Anatase | Given by the authors | Hydrolysis in the presence of dodecylamine followed by annealing | Schertel et al.,[6] |
| 0.5 µm (≲ 5%) | Carbon-doped rutile | Given by the authors | Hydrolysis in the presence of dodecylamine followed by annealing in Ar atmosphere to produce C-doped rutile | Moon et al., [7] |
| 0.4-0.6 µm (≲ 5%) | Amorphous | Given by the authors | Hydrolysis in the presence of dodecylamine | Tanaka et al.,[8] |
| 0.3 µm (≲ 5%) | Amorphous | Given by the authors | Hydrolysis in the presence of KCl | Eiden-Assmann et al.,[9] |
| 0.8 µm (≲ 5%) | Amorphous | Given the authors | Hydrolysis in the presence of diblock co-polymer Lutensol (porous and not clear if morphology after annealing is preserved) | Eiden-Assmann et al.,[9] |
| 0.3-1 µm (5-10%) | Anatase | Given by the authors | Solvothermal synthesis in the presence of EG and acetylacetone | Alam et al.,[10] |
| 0.8 µm (7%) | Amorphous | Given by the authors | Hydrolysis in the presence of NaCl | Eiden-Assmann et al.,[9] |
| 0.85 µm (7%) | Anatase or amorphous | From SEM images (average matches that given by author) | Non-aqueous solvothermal (transient product toward a yolk-shell structure so not characterized) | Li et al.,[11] |
| 0.4 µm (7%) | Amorphous | From TEM images (the most uniform) | Hydrolysis with water | Kojima et al.,[12] |
| 0.7 µm (6%); 1.3 µm (14%); 2.3 µm (25%) | Amorphous | Given by the authors in SI | Hydrolysis over titanium isopropoxide previously treated with carboxylic acids at 90 C. | Tsai et al.,[13] |
| 0.4-0.5 µm (10%) | Rutile | Given by the authors | Hydrolysis with water followed by annealing | Hershey et al.,[14] |
| 0.2-1 µm (~10%) | Anatase | Assuming the contraction in size given by the authors (~25%) | Hydrolysis with NaOH at low temperature (-5 °C) followed by annealing | Poluboyarinov et al.,[15] |
| 0.4-0.6 µm (~10%) | Amorphous | Given by the authors | Hydrolysis with water | Jean et al.,[16] |
| 0.3 µm (~10%) | Amorphous | Given by the authors | Hydrolysis with water in the presence of hydroxypropyl cellulose | Jean et al.,[17] |
| 0.9 µm (~15%), 1.7 (~15%) | Rutile | Lorentzian/Gauss simulations of size distribution images | Acid hydrolysis at 95 °C of $TiCl_4$ (extremely slow reaction time, from 2-8 weeks depending on size) | Matijevic at al.,[18] |
| 0.6 µm (20%) | Amorphous | Given by the authors | Hydrolysis in the presence of Diblock co-polymer-copolymer Pluronic | Eiden-Assmann et al.,[9] |
| 0.7 µm (~15%), 3 µm (~15%) | Carbon-doped anatase | From SEM images main text and SI | Non-aqueous solvothermal method followed by carbonate doping and annealing (highly mesoporous microspheres, $n_{eff} < 2$)) | Liu et al.,[19] |
| 1.1 µm (~15%) | Amorphous | From TEM images | Hydrolysis with water in absence of shear | Look and Zukoski,[20] |
| 0.4 (~15%)-0.8 µm (~25%) | Amorphous | Given by the authors | Hydrolysis in the presence of $NH_3$ (aq.) | Tanaka et al.,[8] |
| 0.2-0.6 µm (~20%) | Amorphous | Given by the authors | Hydrolysis with water (pioneering work) | Barringer et al.,[21,22] |

**Table S2:** Experimental conditions for synthesizing amorphous $TiO_2$ particles of various sizes. The time for visual onset of turbidity (τ) is shown to provide insight and for the sake of reproducibility. In all cases, the reaction proceeded for an additional 3 hours following the appearance of turbidity. The Ti concentrations used here, normalized to acetone volume, are substantially higher than those reported previously, increasing from 0.68-1.2 mM to 15-40 mM in the present study.[1] The data indicate that the nucleation rate plays a key role in determining the average size of $TiO_2$ particles. Taking as reference the conditions optimized to produce amorphous particles with a diameter of 1.1 µm, increasing the amount of the coordinating cosolvent, EG, by ≈ 25 % results in a size increase to 1.6 µm. In contrast, increasing the $NH_3$(aq)/Ti (and so the $H_2O$/Ti ratio as $NH_3$ is supplied as an aqueous solution) by a factor of four reduces the particle size from 1.1 to 0.8 micrometers.

| **Size (µm) and SD (%)** | | **Acetone (mL)** | **EG (mL)** | **Ti butoxide (µL)** | **$NH_3$ (aq) (25wt.%) (µL)** | **τ (min)** |
|---|---|---|---|---|---|---|
| After hydrolysis | Anatase | | | | | |
| 0.8 (5%) | 0.5 (5%) | 54 | 6.0 | 300 | 150 | ≈ 1 |
| 1.1 (4%) | 0.8 (5%) | 54 | 6.0 | 450 | 60 | ≈ 1 |
| 1.6 (5%) | 1.1 (4%) | 54 | 7.5 | 450 | 60 | ≈ 5 |
| 2.4 (4%) | 1.6 (4%) | 54 | 10.5 | 600 | 75 | ≈ 30 |
| 3.0 (5%) | 2.1 (5%) | 54 | 10.5 | 750 | 45 | ≈ 30 |

**Figure S1:** TEM images of particles hydrolyzed using $H_2O$ and $NH_3$. When $H_2O$ is used as the hydrolyzing agent instead of $NH_3$, pronounced particle aggregation occurs, supporting the hypothesis that $NH_3$ also acts as an electrostatic stabilizer. This stabilizing effect is essential for achieving uniform particles at large sizes, as observed in the $NH_3$-hydrolyzed samples. If particle-particle aggregation occurs, the diffusion of polymeric species through the liquid phase toward individual particle surfaces is no longer possible. As a result, the formation of large and uniform particles becomes unfeasible. For the particles shown in the figure, the volumes of acetone, ethylene glycol (EG), titanium butoxide, and either $H_2O$ or $NH_3$ were held constant at 54 mL, 6 mL, 300 µL, and 15 µL, respectively. The hydrolysis process was conducted for 3 hours.

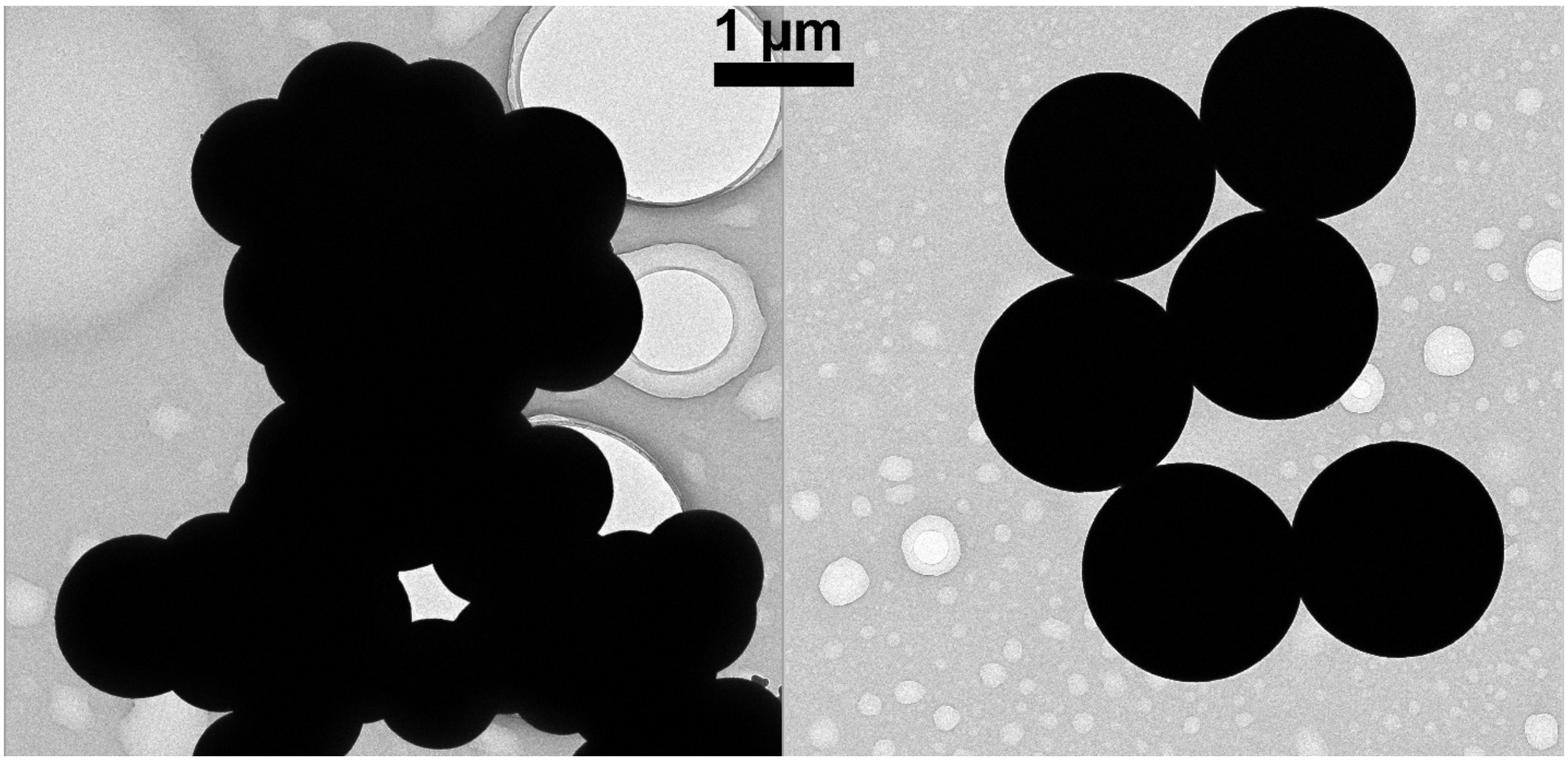

**Figure S2:** SEM image corresponding to the transformation of amorphous particles with an average size of 2.4 μm into anatase particles through thermal annealing at 500 °C. Even though the thermal treatment led to a size contraction of about 30% (from 2.4 to 1.6), the size uniformity of the particles was preserved after crystallization to the anatase phase. However, an as mentioned in the manuscript, during the course of this investigation, we found that the resulting anatases became unprocessable. This behavior probably originates from incipient sintering and/or the development of hydrophobic surfaces during the heating at 500 °C.

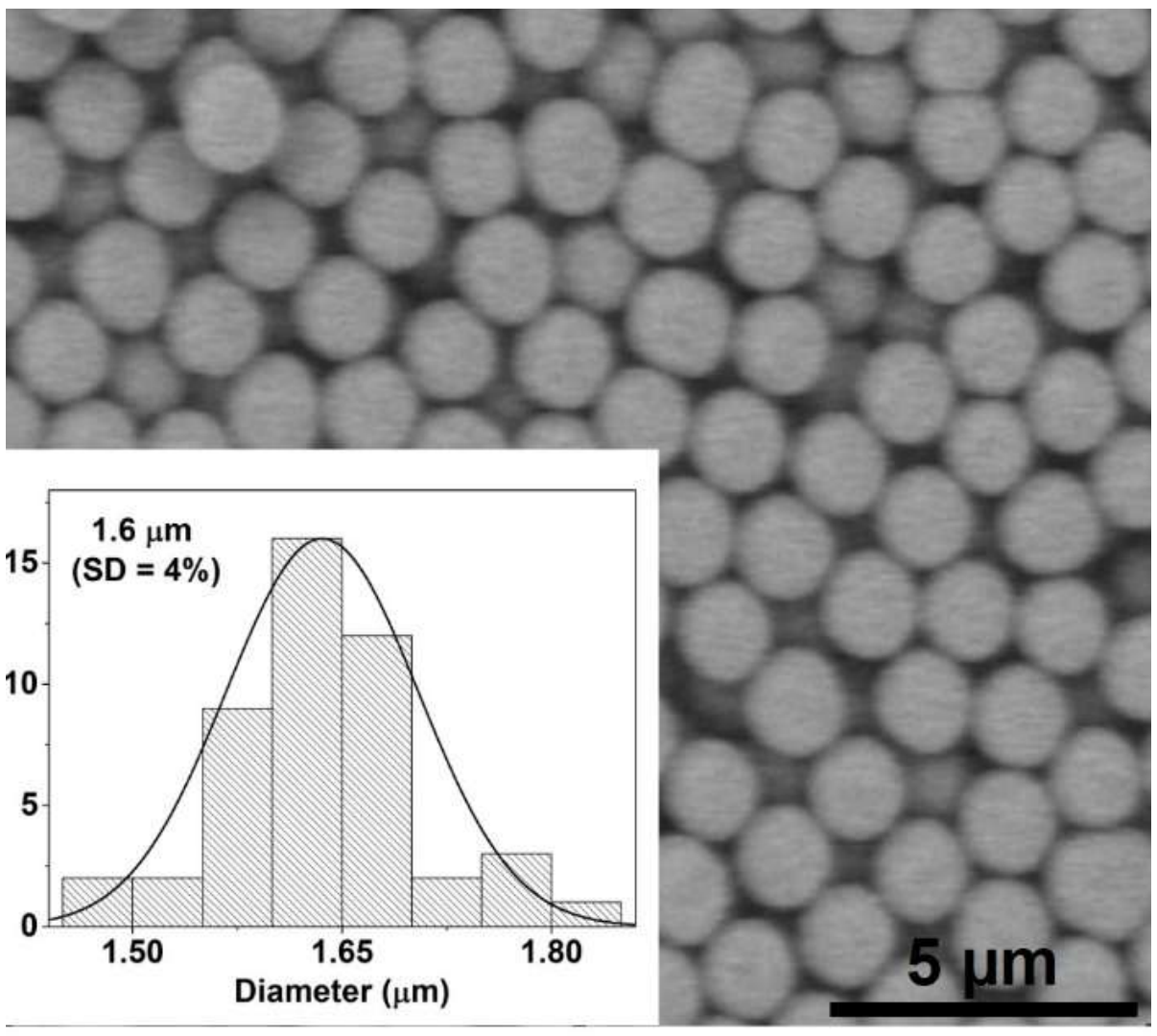

**Figure S3:** X-ray diffraction pattern (XRD) and IR absorption spectrum after annealing at 350 °C/10h an amorphous sample treated with UV irradiation. The XRD pattern and the IR spectrum correspond to anatase.

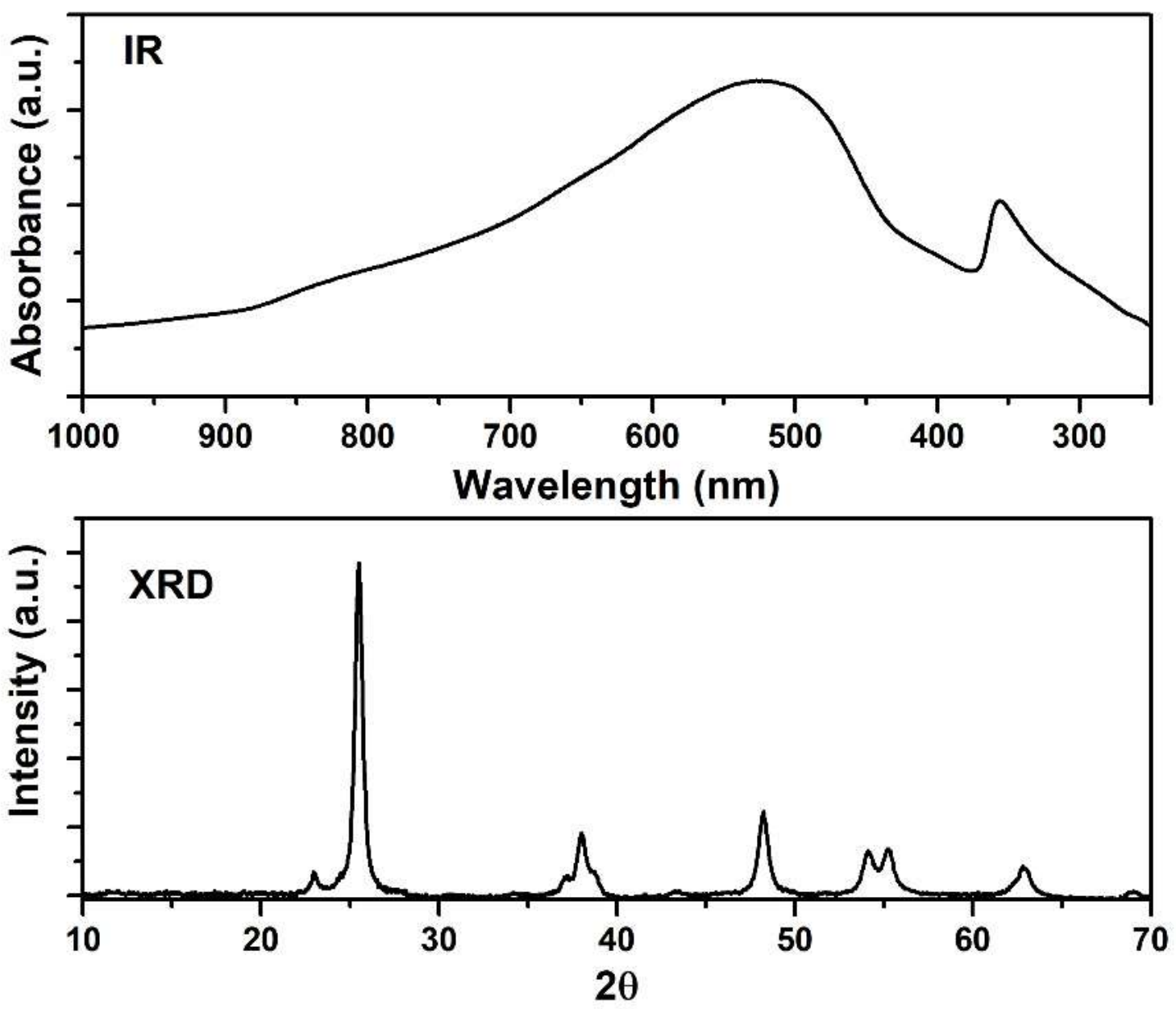

**Figure S4:** (a) IR and (b) DTA/TG curves of freshly prepared samples before and after UV treatment. DT in panel (b) stands for temperature difference. The IR spectra in (a) clearly show that the characteristic bands of CH (2950–2850 $cm^{-1}$) and C–O stretching (1100–1000 $cm^{-1}$), attributed to glycolate and ethylene glycol units,[23] disappear following UV irradiation. This spectral change was supported by elemental analysis, which revealed a substantial drop in residual carbon content from 15 wt.% to approximately 2–3 wt.%. The content of elemental H remained high (2–3 wt.%), which suggests that the UV treatment also generated some hydroxylation on the surface (chemisorbed and physisorbed $H_2O$). This result aligns with previous findings on UV-treated $TiO_2$ films,[24] and with the thermal behavior shown in (b), where a significant portion of weight loss in the UV-treated sample occurs below 200 °C. Prior to treatment, the total mass loss is about 40 wt.%, with over 30 wt.% attributable to organic matter (mainly associated with the two exothermic peaks in the DTA). After UV exposure, the exothermic peak at 500 °C vanishes, and the overall weight loss decreases to approximately 25 wt.%, with organic matter accounting for less than ~10 wt.%. Importantly, as noted in the main manuscript, the pristine amorphous particles are chemically unstable and prone to degradation during prolonged storage under laboratory conditions. Although the UV treatment led to a similar outcome, the exact values shown in this characterization may vary.

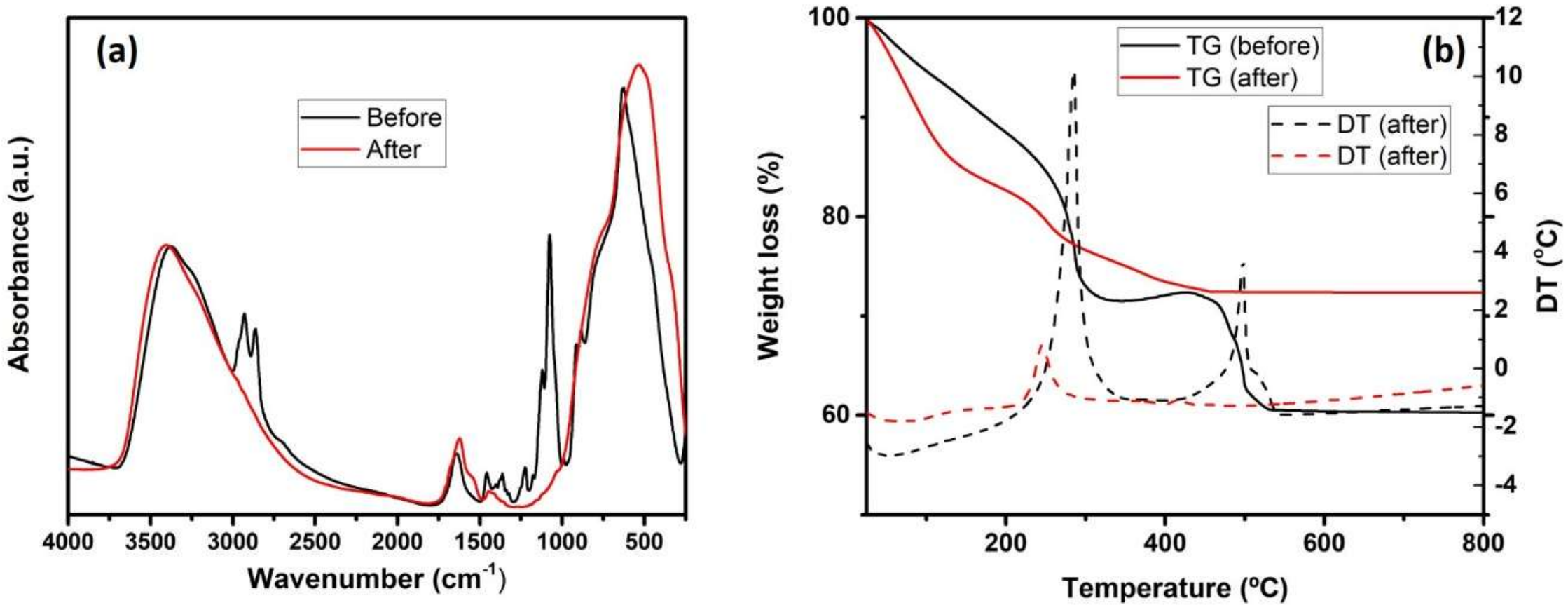

**Figure S5:** SEM image of a specific formulation in which the hydrolysis product was washed with water. In this case, secondary nucleation occurring during the washing stage is clearly distinguishable from the uniform particles obtained from the primary nucleation and growth processes. Such separation would not be observable if the resulting particle sizes were similar, and any overlap, regardless of the experimental conditions, necessarily results in contamination. Notably, some formulations can be washed with water without any evidence of secondary nucleation, which helps explain why other authors have recommended water as a washing agent.

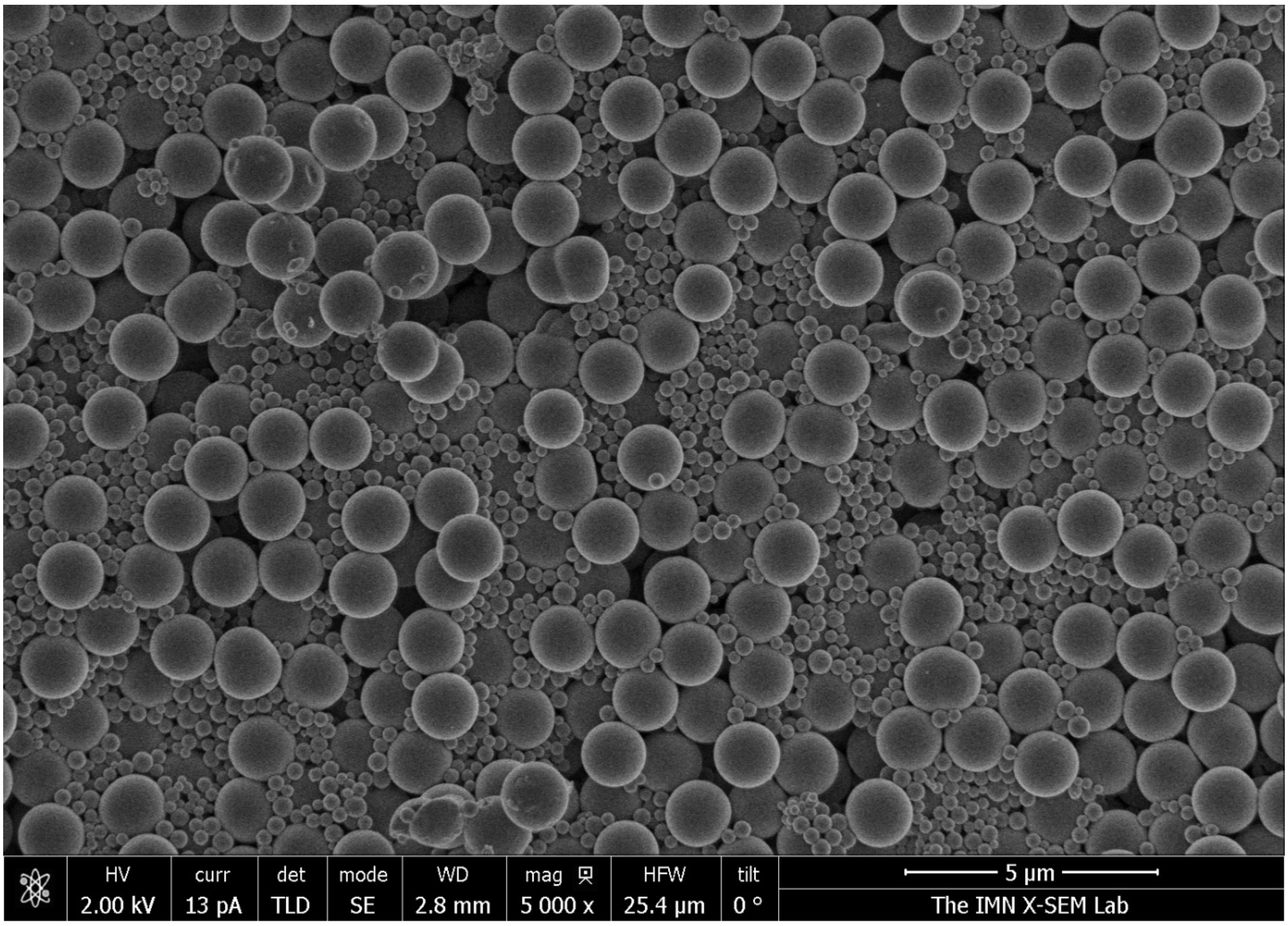

## List of references supporting information